\documentclass{nature}

\usepackage{color}
\usepackage{url}
\usepackage{graphicx}
\usepackage{aas_macros}
\usepackage{amssymb,amsmath}
\usepackage{multirow}

\def\lsim{~\rlap{$<$}{\lower 1.0ex\hbox{$\sim$}}}
\def\gsim{~\rlap{$>$}{\lower 1.0ex\hbox{$\sim$}}}
\def\Mdot{\dot{M}}
\def\Msun{ \rm M_{ \odot } }

\newcommand{\mnrasl}{MNRAS Lett.}

\renewcommand{\thefootnote}{\alph{footnote}}

\long\def\symbolfootnote[#1]#2{\begingroup%
\def\thefootnote{\fnsymbol{footnote}}\footnote[#1]{#2}\endgroup}

\title{Relativistic boost as the cause of periodicity in a massive black-hole binary candidate}

\author{Daniel J. D'Orazio$^{1}$, Zolt\'an Haiman$^{1}$ \& David Schiminovich$^1$}

\begin{document}

\maketitle

\noindent To appear as a {\sl Letter} in the September 17, 2015 isse of {\sl Nature} 

\begin{affiliations}
\item  Department of Astronomy, Columbia University, 550 West 120th Street, New York, NY 10027
\end{affiliations}

\begin{abstract}
{ Because most large galaxies contain a central black hole, and galaxies
  often merge\cite{KormendyHo2013}, black-hole binaries are expected
  to be common in galactic nuclei\cite{BBR1980}.  Although they cannot
  be imaged, periodicities in the light curves of quasars have been
  interpreted as evidence for
  binaries\cite{Komossa2006,Valtonen+2008,Liu+2015}, most recently in
  PG~1302-102, with a short rest-frame optical period of
  4~yr\cite{Graham+2015}. If the orbital period matches this value,
  then for the range of estimated black hole masses the components
  would be separated by 0.007-0.017 pc, implying relativistic orbital
  speeds. There has been much debate over whether black hole orbits
  could be smaller than 1 pc\cite{MM03}.  Here we show that the
  amplitude and the sinusoid-like shape of the variability of
  PG~1302-102 can be fit by relativistic Doppler boosting of emission
  from a compact, steadily accreting, unequal-mass binary.  We predict
  that brightness variations in the ultraviolet light curve track
  those in the optical, but with a 2-3 times larger amplitude. This
  prediction is relatively insensitive to the details of the emission
  process, and is consistent with archival UV data.  Follow-up UV and
  optical observations in the next few years can test this prediction
  and confirm the existence of a binary black hole in the relativistic
  regime. }
\end{abstract} 

Assuming PG~1302-102 is a binary, it is natural to attribute its
optical emission to gas that is bound to each black hole, forming
circumprimary and circumsecondary accretion flows. Such flows, forming
``minidisks'', are generically found in high-resolution 2D and 3D
hydrodynamical simulations that include the black holes in their
simulated
domain\cite{Hayasaki+2008,ShiKrolik2012,Roedig+2012,Dorazio+2013,Nixon+2013,Farris+2014,Dunhill+2015,ShiKrolik2015}.
Assuming a circular orbit, the velocity of the lower-mass secondary
black hole is
\begin{equation}
\nonumber
v_2= \left(\frac{2\pi}{1+q}\right) \left(\frac{GM}{4\pi^2P}\right)^{1/3} 
=8,500 
\left(\frac{1.5}{1+q}\right) 
\left(\frac{M}{10^{8.5}{\rm M_\odot}}\right)^{1/3}
\left(\frac{P}{4.04~{\rm yr}}\right)^{-1/3} \,\,\, {\rm km~s^{-1}},
\end{equation}
or $\sim 0.03c$ for the fiducial parameters above, where $M=M_1+M_2$
is the total binary mass, $M_{1,2}$ are the individual masses,
$q=M_2/M_1\leq1$ is the mass ratio, $P$ is the orbital period, and $c$
is the speed of light.  The primary's orbital velocity is $v_1=qv_2$.
Even if a minidisk has a steady intrinsic rest-frame luminosity, its
apparent flux on Earth is modulated by relativistic Doppler beaming.
The photon frequencies suffer relativistic Doppler shift by the factor
$D=[\Gamma(1-\beta_{||})]^{-1}$, where $\Gamma=(1-\beta^2)^{-1/2}$ is
the Lorentz factor, $\beta=v/c$ is the three-dimensional velocity $v$
in units of the speed of light, and $\beta_{||}=\beta\cos\phi\sin i$
is the component of the velocity along the line of sight, with $i$ and
$\phi$ the orbital inclination and phase.  Because the photon
phase-space density $\propto F_\nu / \nu^3$ is invariant in special
relativity, the apparent flux $F_\nu$ at a fixed observed frequency
$\nu$ is modified from the flux of a stationary source $F_\nu^{\rm 0}$
to $F_\nu =D^{3}F^{\rm 0}_{D^{-1}\nu}=D^{3-\alpha}F^{\rm 0}_{\nu}$.
The last step assumes an intrinsic power-law spectrum $F^{\rm
  0}_{\nu}\propto \nu^{\alpha}$.  To first order in $v/c$, this causes
a sinusoidal modulation of the apparent flux along the orbit, by a
fractional amplitude $\Delta F_\nu/F_\nu = \pm
(3-\alpha)(v\cos\phi/c)\sin i$.  Although light-travel time
modulations appear at the same order, they are subdominant to the
Doppler modulation. This modulation is analogous to periodic
modulations from relativistic Doppler boost
predicted\cite{LoebGaudi2003} and observed for extrasolar
planets\cite{Kerkwijk+2010,MH2010} and for a double white dwarf
binary\cite{Shporer+2010}, but here it has a much higher amplitude.

The light-curve of PG~1302-102 is well measured over approximately two
periods ($\approx 10$ years).  The amplitude of the variability is
$\pm$0.14 mag (measured in the optical $V$
band\cite{Djorgovski+2010}), corresponding to $\Delta
F_\nu/F_\nu=\pm0.14$.  The spectrum of PG~1302-102 in and around the
$V$ band is well approximated by a double power-law, with
$\alpha\approx 0.7$ (between $0.50-0.55\mu$m) and $\alpha\approx 1.4$
(between $0.55-0.6\mu$m), apart from small deviations caused by broad
lines.  We obtain an effective single slope $\alpha_{\rm opt}=1.1$
over the entire $V$ band.  We conclude that the 14\% variability can
be attributed to relativistic beaming for a line-of-sight velocity
amplitude of $v\sin i=0.074~c=22,000~{\rm km~s^{-1}}$.

While large, this velocity can be realised for a massive (high $M$)
but unequal-mass (low $q$) binary, whose orbit is viewed not too far
from edge-on (high $\sin i$).  In Fig.~1, we show the required
combination of these three parameters that would produce a 0.14 mag
variability in the sum-total of Doppler-shifted emission from the
primary and the secondary black hole.  As the figure shows, the
required mass is $\gsim 10^{9.1}~{\rm M_\odot}$, consistent with the
high end of the range inferred for PG~1302-102.  The orbital
inclination can be in the range of $i=60-90^\circ$.  The mass ratio
$q$ has to be low $q\lsim 0.3$, which is consistent with expectations
based on cosmological galaxy merger models\cite{Volonteri+2003}, and
also with the identification of the optical and binary periods (for
$q\gsim 0.3$, hydrodynamical simulations would have predicted that the
mass accretion rates fluctuate with a period several times longer than
the orbital period\cite{PG1302-Dan}).

As Fig.~1 shows, fully accounting for the observed optical variability
also requires that the bulk ($f_2\gsim 80\%$) of the optical emission
arises from gas bound to the faster-moving secondary black hole. We
find that this condition is naturally satisfied for unequal-mass black
holes. Hydrodynamical simulations have shown that in the mass ratio
range $0.03\lsim q\lsim 0.1$, the accretion rate onto the secondary is
a factor of $10-20$ higher than onto the primary\cite{Farris+2014}.
Because the secondary captures most of the accreting gas from the
circumbinary disk, the primary is ``starved'', and radiates with a
much lower efficiency.  In the ($M,q$) ranges favoured by the beaming
scenario, we find that the primary contributes less than 1\%, and the
circumbinary disk contributes less than 20\% to the total luminosity,
leaving the secondary as the dominant source of emission in the
three--component system (see details in {\it Methods}).

The optical light curve of PG~1302-102 appears remarkably sinusoidal
compared to the best-studied previous quasi-periodic quasar binary
black hole candidate, which shows periodic bursts\cite{Valtonen+2008}.
Nevertheless, the light curve shape deviates from a pure sinusoid. In
order to see if such deviations naturally arise within our model, we
maximised the Bayesian likelihood over five parameters (period $P$,
velocity amplitude $K$, eccentricity $e$, argument of pericentre
$\omega$, and an arbitrary reference time $t_0$) of a Kepler
orbit\cite{WrightGaudi2012} and fit the observed optical light-curve.
In this procedure, we accounted for additional stochastic physical
variability with a broken power-law power spectrum (i.e. a ``damped
random walk'' \cite{Kelly:2009:DRW}) described by two additional
parameters.  This analysis indeed returns a best-fit with a non-zero
eccentricity of $e ={0.09}^{+0.07}_{-0.06}$, although a Bayesian
criterion does not favor this model over a pure sinusoid with fewer
parameters (see {\it Methods}).  We have considered an alternative
model to explain PG~1302-102's optical variability, in which the
luminosity variations track the fluctuations in the mass accretion
rate predicted in hydrodynamical
simulations\cite{MM2008,ShiKrolik2012,Roedig+2012,Dorazio+2013,Farris+2014}.
However, the amplitude of these hydrodynamic fluctuations are large
(order unity), and their shape is bursty, rather than
sinusoid-like\cite{Dorazio+2013,Farris+2014,ShiKrolik2015}; as a
result, we find that they provide a poorer fit to the observations
(see Fig.~2 and {\it Methods}).  Furthermore, for mass ratios $q\gsim
0.05$, hydrodynamical simulations predict a characteristic pattern of
periodicities at multiple frequencies, but an analysis of the
periodogram of PG~1302-102 has not uncovered evidence for multiple
peaks\cite{PG1302-Maria}.

A simple observational test of relativistic beaming is possible due to
the strong frequency-dependence of PG~1302-102's spectral slope
$\alpha=d\ln F_\nu /d\ln\nu$.  PG~1302-102's continuum spectrum is
nearly flat with a slope $\beta_{\rm FUV}\equiv d\ln F_\lambda
/d\ln\lambda=0$ in the far UV (0.145-0.1525$\mu$m) band, and shows a
tilt with $\beta_{\rm NUV}=-0.95$ in the 0.20-0.26$\mu$m near UV range
(see Fig.~3 and {\it Methods}).  These translate to $\alpha_{\rm
  FUV}=-2$ and $\alpha_{\rm NUV}=-1.05$ in these bands, compared to
the value $\alpha=1.1$ in the optical.  The UV emission can be
attributed to the same minidisks responsible for the optical light,
and would therefore share the same Doppler shifts in frequency. These
Doppler shifts would translate into UV variability that is larger by a
factor of $(3-\alpha)_{\rm FUV}/(3-\alpha)_{\rm opt}=5/1.9 = 2.63$ and
$(3-\alpha)_{\rm NUV}/(3-\alpha)_{\rm opt}=4.05/1.9 = 2.13$ compared
to the optical, and reach the maximum amplitudes of $\pm 37\%$ (FUV)
and 30\% (NUV).

PG~1302-102 has five separate UV spectra dated between 1992 and 2011,
taken with instruments on the Hubble Space Telescope (HST) and on the
GALEX satellite (see Fig.~3), as well as additional photometric
observations with GALEX at 4 different times between 2006 and 2009
(shown in Fig.~2).  The brightness variations in both the far-- and
near--UV bands show variability resembling the optical variability,
but with a larger amplitude.  Adopting the parameters of our best--fit
sinusoid model, and allowing only the amplitude to vary, we find that
the UV data yields the best--fit variability amplitudes of $\Delta
F_\nu/F_\nu\mid_{\rm FUV} = \pm(35.0 \pm3.9)\%$ and $\Delta
F_\nu/F_\nu\mid_{\rm NUV} = \pm(29.5 \pm2.4)\%$ (shown in
Fig. 2). These amplitudes are factors of $(2.57\pm0.28)$ and
$(2.17\pm0.17)$ higher than in the optical, in excellent agreement
with the values 2.63 and 2.13 expected from the corresponding spectral
slopes.

Relativistic beaming provides a simple and robust explanation of
PG~1302-102's optical periodicity.  The prediction that the larger UV
variations should track the optical light-curve can be tested
rigorously in the future with measurements of the optical and UV
brightness at or near the same time, repeated two or more times,
separated by a few months to $\sim 2$ years, covering up to half of
the optical period. A positive result will constitute the first
detection of relativistic massive black hole binary motion; it will
also serve as a confirmation of the binary nature of PG~1302-102,
remove the ambiguity in the orbital period, and tightly constrain the
binary parameters to be close to those shown in Fig.~1.

\vspace{35pt}

\clearpage
\newpage


\begin{addendum}
 \item The authors thank Matthew Graham, Jules Halpern, Adrian
   Price-Whelan, Jeff Andrews, Maria Charisi, and Eliot Quataert for
   useful discussions. We also thank M. Graham for providing the
   optical data in electronic form. This work was supported by the
   National Science Foundation Graduate Research Fellowship under
   Grant No. DGE1144155 (D.J.D.) and by NASA grant NNX11AE05G (to Z.H.).

\vspace{-0.2\baselineskip}
\item[Contributions] Z.H. conceived and supervised the project,
  performed the orbital velocity calculations, and wrote the first
  draft of the paper. D.J.D. computed the emission models and
  performed the fits to the observed light-curve. D.S.  analysed the
  archival UV data. All authors contributed to the text.
 
\vspace{-0.2\baselineskip}
 \item[Competing Interests] The authors declare that they have no
   competing financial interests.

\vspace{-0.2\baselineskip}
 \item[Correspondence] Correspondence and requests for materials
   should be addressed to Z.H. \\(email: zoltan@astro.columbia.edu).

\end{addendum}

\clearpage
\newpage

\begin{figure}
\centering
\includegraphics[width=0.9\textwidth]{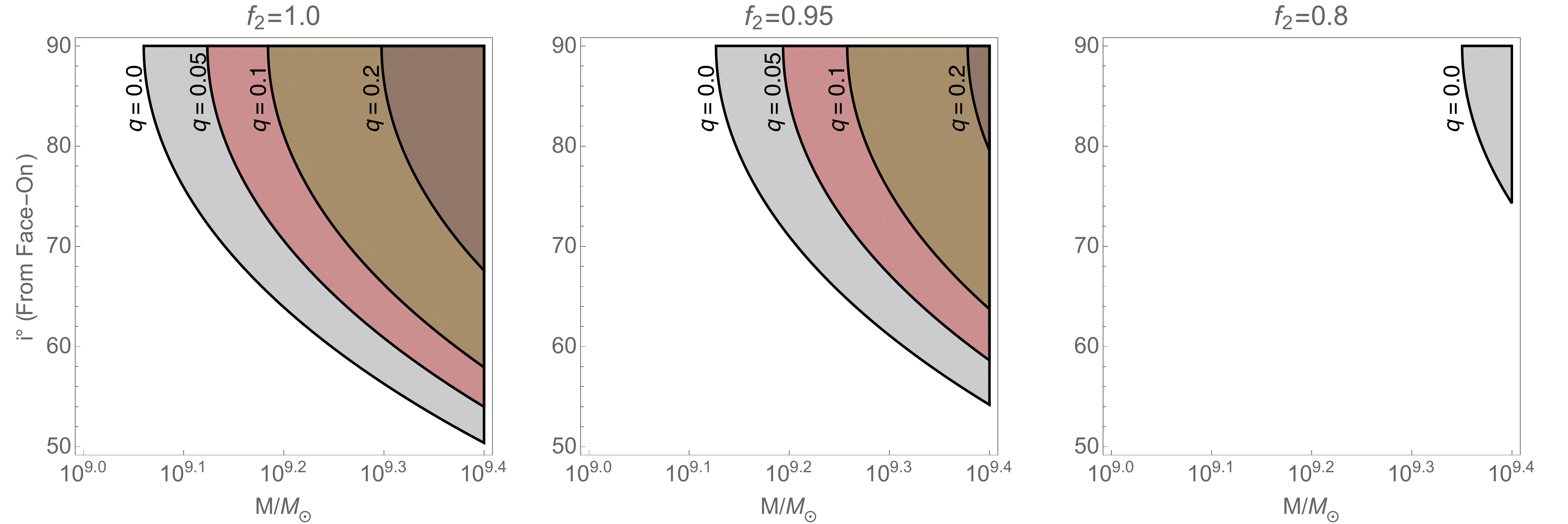}
\caption{{\bf $\mid$ Binary parameters producing the optical flux
    variations of PG~1302-102 by relativistic boost.}  Shaded regions
  mark combinations of binary mass $M$, mass ratio $q=M_2/M_1$, and
  inclination $i$ causing $>$13.5\% flux variability (or line-of-sight
  velocity amplitude $(v/c)\sin i\geq 0.07$), computed from the
  Doppler factor $D^{3-\alpha}$ with the effective spectral slope of
  $\alpha=1.1$ in the $V$ band, including emission from the primary,
  as well as from the secondary black hole.  The three panels assume
  fractions $f_2=1.0, 0.95$, or $0.8$ of the total luminosity arising
  from the secondary black hole; these values are consistent with
  fractions found in hydrodynamical simulations\cite{Farris+2014} (see
  {\it Methods}).}
\label{fig:Msini}
\end{figure}

\clearpage
\newpage

\begin{figure}
\centering
\includegraphics[width=0.95\textwidth]{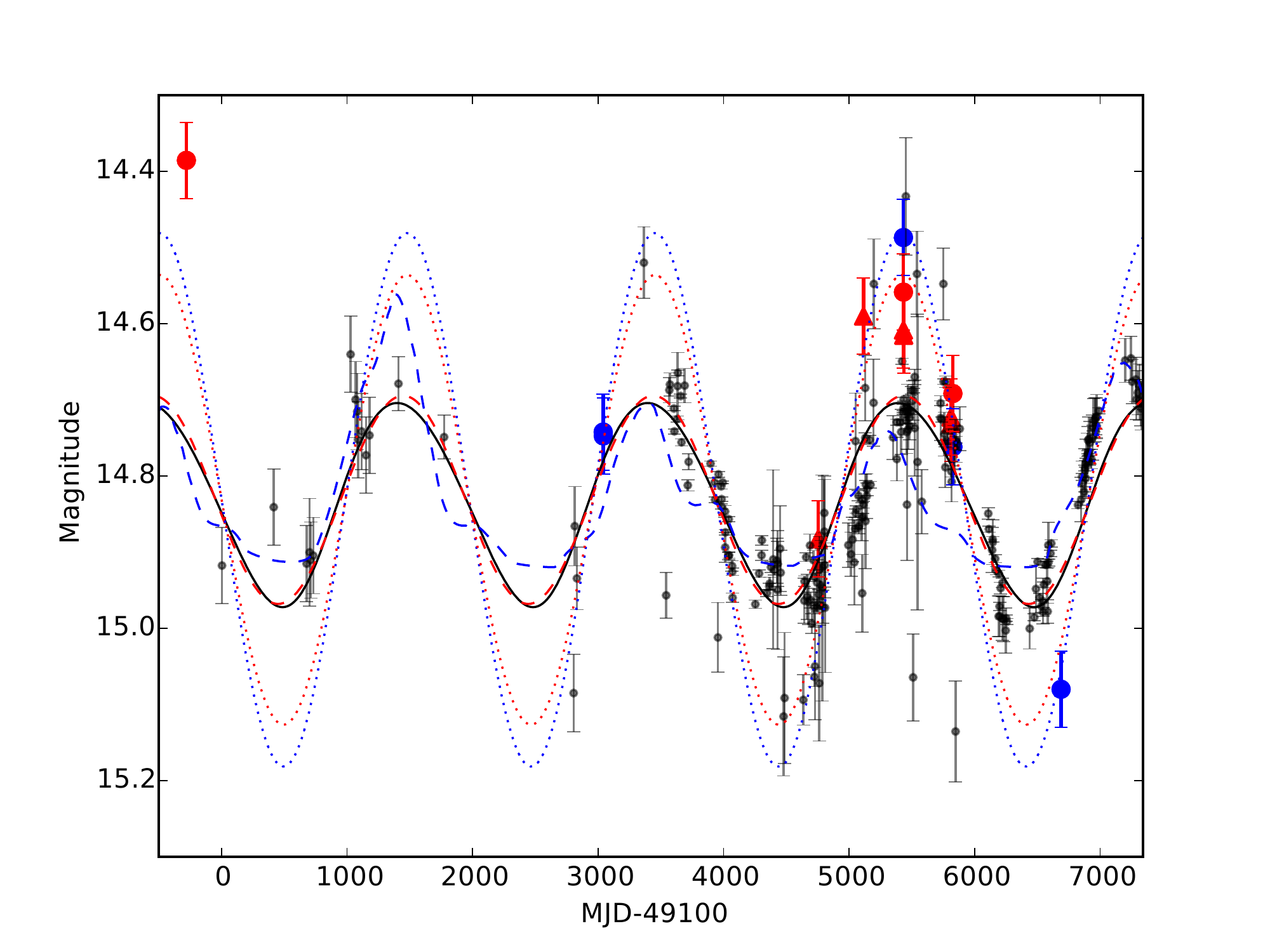}
\vspace{-15\baselineskip}
\caption{{\bf $\mid$ The optical and UV light-curve of PG~1302-102.}
  Black points with $1\sigma$ errors are optical
  data\cite{Graham+2015}, superimposed with a best-fitting sinusoid
  (red dashed curve). The solid black curve shows the best--fit
  relativistic light--curve. The blue dashed curve shows the best-fit
  model obtained by scaling the mass-accretion rate found in a
  hydrodynamical simulation of an unequal-mass ($q=0.1$)
  binary\cite{Dorazio+2013}.  The additional circular data points with
  $1\sigma$ errors show archival near-UV (red) and far-UV (blue)
  spectral observations; the red triangles show archival photometric
  near-UV data-points (see Fig.~3).  The UV data include an arbitrary
  overall normalisation to match the mean optical brightness.  The
  dotted red and blue curves show the best-fit relativistic optical
  light curve with amplitude scaled up by factors of 2.17 and 2.57,
  which best match the NUV and FUV data, respectively.}
\label{fig:lightcurve}
\end{figure}

\clearpage
\newpage

\begin{figure}
\centering
\vspace{-\baselineskip}
\includegraphics[width=0.75\textwidth]{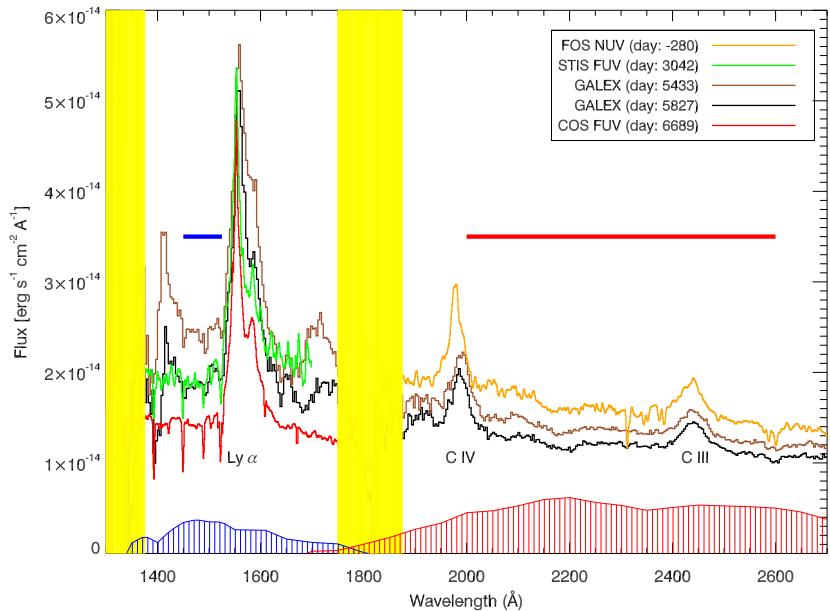}
\vspace{-\baselineskip}
\caption{{\bf $\mid$ Archival ultraviolet spectra of PG~1302-102 from
    1992-2011.}  Far- and near-UV spectra obtained by the FOS and STIS
  instruments on the Hubble Space Telescope (HST) and by GALEX are
  shown. Dates are in MJD (modified Julian day)-49100. Vertical yellow
  bands mark regions outside the spectroscopic range of both GALEX and
  HST and contain no useful spectral data.  From each spectrum,
  average flux measurements were computed in one or both of the two UV
  bands (shown in Fig.~2).  GALEX photometric band shapes for FUV and
  NUV photometry are shown for reference as shaded blue and red
  curves, respectively.  Additional GALEX NUV photometric data were
  also used in Fig.~2.  The UV spectra show an offset by as as much as
  $\pm$30\%, close to the value expected from relativistic boost (see
  {\it Methods}).}
\label{fig:spectra}
\end{figure}


\section*{\large\textcolor{blue}{METHODS}}
\subsection{V-band emission from a three--component system in PG~1302-102.}

Here we assume that the PG~1302-102 supermassive black hole (SMBH)
binary system includes three distinct luminous components: a
circumbinary disk (CBD), as well as an actively accreting primary and
secondary SMBH.  The optical brightness of each of the three
components can be estimated once their accretion rates and the BH
masses $M_1$ and $M_2$ are specified.  Using the absolute V-band
magnitude of PG 1302-102, $M_V = -25.81$ and applying a bolometric
correction $BC \approx 10$, \cite{RichardsQBCs:2006} we infer the
total bolometric luminosity of $L_{\rm bol}=6.5 (BC/10) \times
10^{46}\, {\rm erg~s^{-1}}$.  Bright quasars with the most massive
SMBHs ($M\gsim 10^9{\rm M_\odot}$), have a typical radiative
efficiency of $\epsilon=0.3$ \cite{YuTremaine2002}.  Adopting this
value, the implied accretion rate is $\Mdot_{\rm tot} =
L_{\rm{bol}}/(\epsilon c^2) = 3.7\, {\rm M_\odot~yr^{-1}}$.

We identify this as the total accretion rate through the CBD, and
require that at small radii, the rate is split between the two black
holes $\Mdot_{\rm tot} = \Mdot_2 + \Mdot_1$ with the ratio $\eta\equiv
\Mdot_2/\Mdot_1$.  Hydrodynamical simulations\cite{Farris+2014} have
found that the secondary captures the large majority of the gas, with
$10\lsim \eta \lsim 20$ for $0.03\lsim q \lsim 0.1$ (where $q\equiv
M_2/M_1$).  Defining the Eddington ratio of the $i^{\rm{th}}$ disk as
its accretion rate scaled by its Eddington--limited rate
$f_{i,\rm{Edd}}\equiv \Mdot_i/\Mdot_{i,\rm{Edd}}$ with $\Mdot_{\rm
  Edd} \equiv L_{\rm Edd}/0.1c^2$ (here $L_{\rm Edd}$ is the Eddington
luminosity for the $i^{\rm th}$ BH, and we have adopted the fiducial
radiative efficiency of 0.1 to be consistent with the standard
definition in the literature), we have
\begin{align}
f_{\rm{CBD,Edd}} & \approx   0.068 \left( \frac{M_{\rm tot}}{10^{9.4} \Msun} \right)^{-1}, \nonumber \\
f_{\rm{1,Edd}} &= f_{\rm{CBD,Edd}}  \frac{ \left( 1+q \right)}{1 + \eta}  \sim 0.0034 \left( \frac{ f_{\rm{CBD,Edd}}}{0.068} \right) \left(\frac{1+q}{1.05}\right) \left( \frac{21}{1+\eta}\right), \nonumber \\
f_{\rm{2,Edd}} &=\eta \frac{f_{\rm{1,Edd}}}{q} \simeq 1.37  \left( \frac{ f_{\rm{1,Edd}}}{0.0034} \right) \left( \frac{\eta}{20}\right) \left( \frac{0.05}{q}\right),
\label{Eq:TDfEdds}
\end{align}
where the subscripts $1$ and $2$ refer to the primary and the
secondary, and $M_{\rm tot} \equiv M_1 + M_2$.  We adopt a standard,
radiatively efficient, geometrically thin, optically thick
Shakura-Sunyaev (SS) disk model\cite{SS73} to compute the luminosities
produced in the CBD and the circum-secondary disk (CSD). Although the
secondary is accreting at a modestly super-Eddington rate, recent 3D
radiation magneto-hydrodynamic simulations of super-Eddington
accretion find radiative efficiencies comparable to the values in
standard thin disk models\cite{Jiang+2014}.  On the other hand, the
primary is accreting below the critical rate $\dot{M}_{\rm{1}}
\lesssim \dot{M}_{\rm{ADAF}} \approx 0.027 (\alpha/0.3)^2
\dot{M}_{\rm{Edd}}$ (with $\alpha$ the viscosity parameter) for which
advection dominates the energy balance\cite{NarayanMcClintock2008}. We
therefore estimate its luminosity from a radiatively inefficient
advection-dominated accretion flow (ADAF)\cite{Mahadevan:1997,
  NMQ:ADAF:1998}, rather than a SS disk.  This interpretation is
supported by the fact that PG~1302-102 is known to be an extended radio
source, with evidence for a jet and bends in the extended radio
structure\cite{Hutchings+2014}, features that are commonly associated
with sub-Eddington sources\cite{WangHoStaubert2003}.

For the radiatively efficient CBD and CSD, the frequency-dependent
luminosity is given by integrating the local modified blackbody flux
over the area of the disk
\begin{align}
L_{\nu} =& 2 \pi \int^{R_{\rm out}}_{R_{\rm in}}{F_{\nu}[T_p(r)] \ r dr} \nonumber \\
F_{\nu} =& \pi \frac{2 \epsilon^{1/2}_{\nu}}{1 + \epsilon^{1/2}_{\nu} } B_{\nu} \qquad \epsilon_{\nu} = \frac{  \kappa^{\rm abs}_{\nu}  }{  \kappa^{\rm abs}_{\nu} +  \kappa^{\rm es}}
\label{Eq:LandF}
\end{align}
where $B_{\nu}$ is the Planck function, $ \kappa^{\rm abs}_{\nu}$ is
the frequency-dependent absorption opacity, and $\kappa^{\rm es}$ is
the electron scattering opacity. We compute the radial disk
photosphere temperature profile $T_p$ by equating the viscous heating
rate to the modified blackbody flux
\begin{align}
\label{Eq:Tp}
& \left[ \frac{3GM\dot{M}}{8 \pi r^3 }  \left[ 1 -  \left(\frac{r_{\rm{ISCO}}}{r}\right)^{1/2}\right]  \right] = \zeta(\nu, T_p) \sigma T^4_p  \\ \nonumber
& \zeta(\nu, T_p) = \frac{15}{\pi^4} \int{   \frac{2 \epsilon^{1/2}_{\nu}(x)}{1 + \epsilon^{1/2}_{\nu}(x) }  \frac{x^3 e^{-x}}{1- e^{-x}}  \ dx }  \quad x \equiv \frac{h \nu}{k_B T_p}.
\end{align}
where $\sigma$ is the Stefan-Boltzmann constant and $k_B$ is
Boltzmann's constant. In solving for the photosphere temperature we
work in the limit that $\kappa^{\rm abs}_{\nu} \ll \kappa^{\rm es}$
following Appendix A of ref.~\cite{TanakaMenou:2010}, and we adopt
$r^i_{\rm{ISCO}}=6GM^i/c^2$ and $(R_{\rm in}, R_{\rm out}) =
(2a,200a), (r^s_{\rm{ISCO}}, a (q/3)^{1/3})$ for the inner and outer
radii of the CBD and CSD, respectively. Here the superscript $i$
refers to the $i^{\rm{th}}$ disk, $a$ is the binary separation, $6GM$
is the location of the innermost stable circular orbit for a
Schwarzshild black hole (our results are insensitive to this choise)
and $a(q/3)^{1/3}$ is the secondary's Hill radius (which provides an
upper limit on the size of the CSD\cite{AL94}).

The optical luminosity of an ADAF is sensitive to the assumed
microphysical parameters and its computation is more complicated than
for a thin disk.  Here we first compute a reference thin-disk
luminosity $L_{\rm SS}$ for the primary, and multiply it by the ratio
of the bolometric luminosity of an ADAF to an equivalent thin-disk
luminosity from ref.~\cite{Mahadevan:1997},
\begin{align}
\frac{L_{ADAF}}{L_{SS}} \sim 0.008  \left( \frac{ \dot{M} / \dot{M}_{\rm{Edd}}}{0.0034}\right)  \left( \frac{\alpha}{0.3} \right)^{-2}.
\label{Eq:LADAF}
\end{align}
For calculating the reference $L_{\rm SS}$, we adopted parameters
consistent with ref.~\cite{Mahadevan:1997}, in particular
$\epsilon=0.1$.  Although the above ratio is for the bolometric
luminosities, we find that it agrees well with the factor of $100$
difference in the V band shown in Figure~6 of
ref. \cite{NMQ:ADAF:1998} between ADAF and thin disk spectra with
parameters similar to PG 1302-102 ($10^9 \Msun$, $\Mdot =
\dot{M}_{\rm{ADAF}}=10^{-1.5} \Mdot_{\rm{Edd}}$, $\alpha \approx
0.3$).

Extended Data Fig.~1 
shows the thin-disk CBD and CSD spectra
for a total Eddington ratio of $f_{\rm{CBD,Edd}}=0.07$, consistent
with the high--mass estimates for PG 1302-102 needed for the beaming
scenario ($M=10^{9.4}~{\rm M_\odot}$ and $q=0.05$).  The red dot shows
the reduced V-band luminosity of an ADAF onto the primary. The
secondary clearly dominates the total V band luminosity, with the
primary contributing less than 1\%, and the CBD contributing $\sim
14\%$.  In practice, the contribution from the CBD becomes
non-negligible only for the smallest binary masses and lowest mass
ratios (reaching 20\% for $M<10^{9}\Msun$ and $q<0.025$).

We compute the contributions of each of the three components to the
total luminosity, $L^V_{\rm tot}= L^V_1 + L^V_2 + L^V_{\rm CBD}$, and
the corresponding total fractional modulation amplitude $\Delta
L^V_{\rm tot} / L^V_{\rm tot} = (\Delta L^V_1 + \Delta L^V_2) /
L^V_{\rm tot}$, for each value of the total mass $M$ and mass ratio
$q$.  The primary is assumed to be Doppler-modulated with a line-of
sight velocity $v_1=-q v_2$ while the emission from the CBD is assumed
constant over time ($\Delta L^V_{\rm
  CBD}=0$). Extended Data Fig.~2 
shows regions in the ($M,q,i$) parameter space where the total
luminosity variation due to relativistic beaming exceeds $14\%$.  This
recreates Fig.~1 of the main text, but using the luminosity
contributions computed self-consistently in the above model, rather
than assuming a constant value of $f_2$. Because the secondary is
found to be dominant, the relativistic beaming scenario is consistent
with a wide range of binary parameters.

\subsection{Model fitting to the PG 1302-102 optical light curve.}
We fit models to the observed light curve of PG 1302-102 by maximising the
Bayesian likelihood $\mathcal{L} \propto {\rm
  det}|{\mbox{Cov}^{D}}{\mbox{Cov}^{ph}}|^{-1/2}\exp(-\chi^2/2)$,
where
\begin{align}
\chi^2 \equiv \mathbf{Y^T} \mathbf{(\mbox{Cov})^{-1}}  \mathbf{Y},
\end{align}
and $\mathbf{Y} \equiv \mathbf{O} - \mathbf{M}$ is the difference
vector between the mean flux predicted in a model and the observed
flux at each observation time $t_i$. Here $\mbox{Cov}$ is the
covariance matrix of flux uncertainties, allowing for correlations
between fluxes measured at different $t_i$.  We include two types of
uncertainties: (1) random (uncorrelated) measurement errors,
\begin{align}
 \mbox{Cov}^{ph}_{ij} = 
 \begin{cases} 
      \sigma^2_i & i=j \\
      0 & i \neq j 
   \end{cases}
\end{align}
where $\sigma^2_i$ is the variance in the photometric measurement for the
$i^{th}$ data point (as reported in ref.~\cite{Graham+2015}), and (2)
correlated noise due to intrinsic quasar variability, with covariance
between the $i^{th}$ and $j^{th}$ data points,
\begin{align}
\mbox{Cov}^{D}_{ij} = \sigma^2_{D} \mbox{exp}\left[ \frac{-|t_i - t_j|}{(1 + z) \tau_D}\right].
\end{align}
The parameters $\sigma_D$ and $\tau_D$ determine the amplitude and
rest--frame coherence time of correlated noise described by the damped
random walk (DRW) model\cite{Kelly:2009:DRW}, and the factor of
$(1+z)$ converts $\tau_D$ to the observer's frame where the $t_i$ are
measured.  These parameters enter the normalisation of the Bayesian
likelihood, and this normalisation must therefore be included when
maximising the likelihood over these
parameters\cite{Kozlowski+2010}. The covariance matrix for the total
noise is given by $\mbox{Cov} = \mbox{Cov}^{D} + \mbox{Cov}^{ph}$. We
assume both types of noise are Gaussian, which provides a good
description of observed quasar variability\cite{Andrae+2013}.

We then fit the following four different types of models to the data:
\begin{itemize}
\item {\it Relativistic beaming model} with 5+2=7 model parameters:
  eccentricity, argument of peri-center, amplitude, phase and orbital
  period, plus the two noise parameters $\sigma_D$ and $\tau_D$.
\item {\it Accretion rate model} with 3+2=5 model parameters:
  amplitude, phase and period, plus the two noise parameters.  This
  model assumes that PG~1302-102's light-curve tracks the mass accretion
  rates predicted in hydrodynamical simulations.  For near-equal mass
  binaries, several studies have found that the mass accretion rates
  fluctuate periodically, but they resemble a series of sharp bursts,
  unlike the smoother, sinusoid-like shape of PG~1302-102's light-curve.
  To our knowledge, only three studies to date have simulated
  unequal-mass ($q\leq 0.1$) SMBH
  binaries\cite{Dorazio+2013,Farris+2014,ShiKrolik2015}. The accretion
  rates for these binaries are less bursty; among all of the cases in
  these three studies, the $q=0.075$ and $q=0.1$ binaries in
  ref~\cite{Dorazio+2013} resemble PG~1302-102's light-curve most closely
  (shown in Extended Data Fig.~3).  Here we adopt the published
  accretion curve for $q=0.1$, and perform a fit to PG~1302-102 by
  allowing an arbitrary linear scaling in time and amplitude, as well
  as a shift in phase; this gives us the three free parameters for
  this model.  (We find that the $q=0.075$ case provides a worse fit.)
\item {\it Sinusoid model} with 3+2 parameters: amplitude, phase and
  period, plus the two noise parameters. This model is equivalent, to
  first order in $v/c$, to the beaming model restricted to a circular
  binary orbit.
\item {\it Constant luminosity model} with 2 parameters: for
  reference only -- to quantify how poor the fit is with only an
  amplitude plus the two noise parameters.

\end{itemize}

In each of these models above, we have fixed the mean flux to equal
its value inferred from the optical data.  We have found that allowing
the mean to be an additional free parameter did not change our
results.  The highest maximum likelihood is found for the beaming
model, with best-fit values of ($P=\left[1996\right]^{+29}_{-35}$
days, $K=\left[0.065\right]^{+0.007}_{-0.006} c$, $e
={0.09}^{+0.07}_{-0.06}$,
$\cos\omega=\left[-0.65\right]^{+1.2}_{-0.06}$,
$t_0=\left[718\right]^{+422}_{-34}$ days), where the reference point
$t_0$ is measured from MJD$-49100$.  Uncertainties are computed with
the \textit{emcee} code\cite{DFM:2013}, which implements a Markov
Chain Monte Carlo algorithm, and which we use to sample the 7D
posterior probability of the model given the G15 data. In practice, we
employ 28 individual chains to sample the posterior for 1024 steps
each. Throwing away the first 600 steps (`burning in'), we run for 424
steps and for each parameter, we quote best-fit values corresponding
to the maximum posterior probability, with errors given by the
$85^{\rm th}$ and $15^{\rm th}$ percentile values (marginalized over
the other six parameters) .  The best-fit noise parameters are found
to be $(\sigma_D, \tau_D)=(0.049^{+0.016}_{-0.001}~{\rm mag},
37.6^{+35.2}_{-1.5}~{\rm days})$.  The best-fit model has a reduced
$\chi^2/(N-1-7) \approx 2.1$, where $N=245$ is the number of data
points.

To assess which of the above model is favoured by the data, we use the
Bayesian Information Criterion (BIC), a standard method for comparing
different models, penalising models with a larger number of free
parameters\cite{KassRaftery1995}.  Specifically, BIC$=-2\ln\mathcal{L}
+ k \ln N$, where the first term is evaluated using the best-fit
parameters in each of the models and where $k$ is the number of model
parameters. We find the following differences $\Delta BIC$ between
pairs of models:

\begin{center}
\vspace{-\baselineskip}
$\begin{array} {l l}
   BIC_{\rm Acc} - BIC_{\rm Beam} = 4.0 \,\,\, {\rm (Beaming~model~preferred~over~accretion~model)} \\  
  BIC_{\rm Acc} - BIC_{\rm Sin} = 14.9 \,\,\, {\rm (Sin~model~strongly~preferred~over~accretion~model)} \\ 
  BIC_{\rm Sin} - BIC_{\rm Beam} = -10.9  \,\,\, {\rm (Sin~model~strongly~preferred~over~eccentric~beaming~model)}  \\
  BIC_{\rm Const} - BIC_{\rm Beam} = 11.5   \,\,\, {\rm (Beaming~model~strongly~preferred~over~pure~noise)} \\ 
  BIC_{\rm Const} - BIC_{\rm Sin} = 22.4 \,\,\, {\rm (Sin~model~strongly~preferred~over~pure~noise)}. 
\end{array}$
\end{center} 

\vspace{-\baselineskip} We conclude that a sinusoid, or equivalently
the beaming model restriced to a circular binary, is the preferred
model.  In particular, this model is very strongly favoured over the
best--fit accretion models (see Extended Data Fig.~3), with $\Delta
BIC>14.9$. For the assumed Gaussian distributions, this corresponds to
an approximate likelihood ratio of $\exp(-14.9/2)\approx 5.7\times
10^{-4}$.  Although our best--fit beaming model has a small non-zero
eccentricity, the 7-parameter eccentric model is disfavoured (by
$\Delta BIC=11.5$) over the 5-parameter circular case.

We have conservatively allowed the amplitude of accretion rate
fluctuations to be a free parameter in the accretion models, but we
note that the accretion rate variability measured in hydrodynamic
simulations exhibits large (order unity) deviations from the mean,
even for $0.05<q<0.1$ binaries \cite{Dorazio+2013, Farris+2014,
  ShiKrolik2015}.  In the accretion rate models, an additional
physical mechanism needs to be invoked to damp the fluctuations to the
smaller $\sim 14\%$ amplitude seen in PG 1302-102 (such as a more
significant contribution from the CBD and/or the primary).

\subsection{Disk Precession.}
The lowest BIC model, with a steady accretion rate and a relativistic
boost from a circular orbit, has a reduced $\chi^2=2.1$, indicating
that the relativistic boost model with intrinsic noise does not fully
describe the observed light-curve.  The residuals could be explained
by a lower-amplitude periodic modulation in the mass accretion rate,
which is expected to have a non-sinusoidal shape (i.e. with sharper
peaks and broader troughs, as mentioned
above\cite{Farris+2014}). Alternatively, the minidisks, which we have
implicitly assumed to be co-planar with the binary orbit, could
instead have a significant tilt\cite{Nixon+2013}.

A circum-secondary minidisk that is tilted with respect to the
binary's orbital plane will precess around the binary angular momentum
vector, causing additional photometric variations due to the changing
projected area of the disk on the sky. The precession timescale can be
estimated from the total angular momentum of the secondary disk and
the torque exerted on it by the primary black hole.  The ratio of the precession
period to the binary's orbital period is \cite{DLai2014},
\begin{align}
\frac{P_{\rm prec}}{P_{\rm orb}}=  -\frac{8}{\sqrt{3}} \frac{\sqrt{1+ q }}{  \cos{\delta}},
\end{align}
where we have chosen the outer edge of the minidisk to coincide with
the secondary's Hill sphere $R_{\rm H} = (q/3)^{1/3}a$, for binary
semi-major axis $a$. This choice gives the largest secondary disk and
the shortest precession rates.  The angle $\delta$ between the disk
angular momentum vector and the binary angular momentum vector can
range from $-\pi/2$ to $\pi/2$. For small binary mass ratios,
consistent with the relativistic beaming scenario, the precession can
be as short as $4.8 P_{\rm orb}$, causing variations on a timescale
spanning the current observations of PG~1302-102. The precession
timescale would be longer ($>20 P_{\rm {orb}}$) for a smaller
secondary disk tidally truncated at $0.27q^{0.3}a$
\cite{Roedig+Krolik+Miller2014}, and with a more inclined
($45^{\circ}$) disk.

\subsection{Archival UV data.}

FUV (0.14-0.175$\mu$m) and NUV (0.19-0.27$\mu$m) spectra of PG~1302-102
were obtained by the Hubble Space Telescope (HST) and the Galaxy
Evolution Explorer (GALEX) since 1992.  HST/Faint Object Spectrograph
(FOS) NUV spectra were obtained on July 17, 1992
(pre-COSTAR)\cite{EvansKoratkar2004}.  HST/Space Telescope Imaging
Spectrograph (STIS) FUV spectra were obtained on August 21,
2001\cite{Cooksey+2008}. GALEX FUV and NUV spectra were obtained on
March 8, 2008 and April 6, 2009 and HST/Cosmic Origins Spectrograph
(COS) FUV spectra were obtained on January 28, 2011. All data are
publicly available through the Mikulski Archive for Space Telescopes
(MAST) at {\it archive.stsci.edu}.  All measurements have been
spectrophotometrically calibrated, and binned or smoothed to
1-3\AA\ resolution.  The spectra (shown in Fig.~2 in the main text)
have errors per bin typically less than 2\% and published absolute
photometric accuracies are better than 5\%. 

From each spectrum, average flux measurements (shown in Fig.~2 in the
main text) were obtained in one or both of two discrete bands: FUV
continuum from 0.145-0.1525$\mu$m for FUV (a range chosen to avoid the
Ly$\alpha$ line) and NUV continuum from 0.20-0.26$\mu$m.  For the
GALEX NUV photometric data (also used in Fig.~2) we adopted a small
correction (0.005 mag) for the transformation from the GALEX NUV to
our NUV continuum band.  GALEX FUV photometric data were not used
because of the significant contribution from redshifted
Ly$\alpha$. Note that the broad lines in the UV spectra (in Fig.~3) do
not show a large $\Delta\lambda=(v/c)\lambda\approx 140\AA$ Doppler
shift. This is unsurprising, since the broad line widths
(2,500-4,500~${\rm km~s^{-1}}$) are much smaller than the inferred
relativistic line-of-sight velocities, and are expected to be produced
by gas at larger radii, unrelated to the rapidly orbiting minidisks
producing the featureless thermal continuum emission\cite{PG1302-Dan}.

\vspace{2\baselineskip}


\clearpage
\newpage

\begin{center}
\includegraphics[scale=0.6,angle=-90]{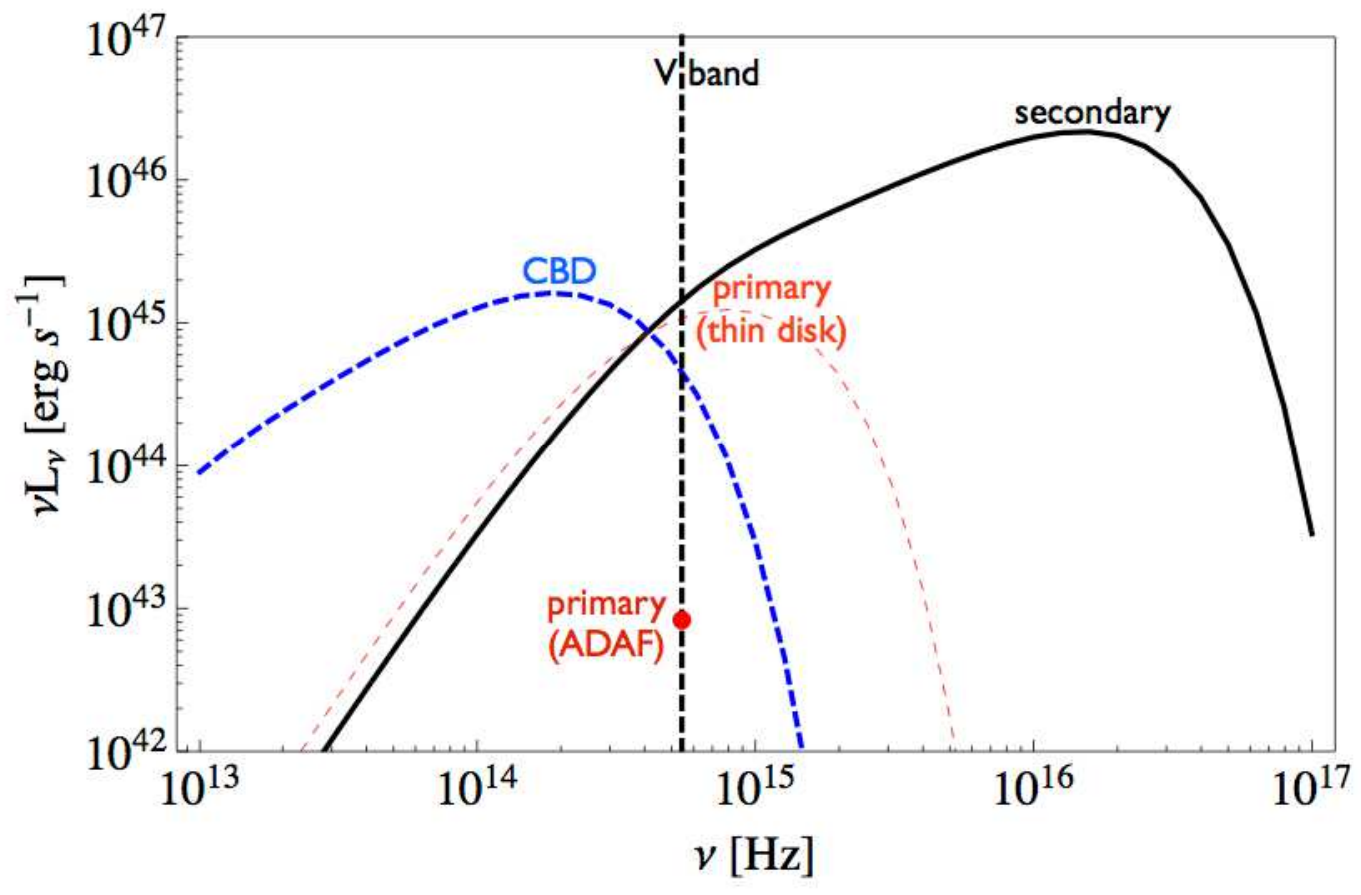} 
\vspace{-1in}
\end{center}
\noindent{\bf Extended Data Figure 1 $\mid$ Model spectrum of PG~1302-102.} Circumbinary
(dashed blue) and circumsecondary (solid black) disk spectra for a
total binary mass of $10^{9.4}$, binary mass ratio $q=0.05$, and ratio
of accretion rates $\dot{M}_2/\dot{M}_1 = 20$.  A vertical dashed line
marks the center of the V-band and the approximate flux from an
advection--dominated accretion flow (ADAF) is shown as a red dot for
the V-band contribution of the primary.  The spectrum for a
radiatively efficient, thin disk around the primary is shown by the
thin red dashed curve for reference.

\clearpage
\newpage

\begin{center}
\resizebox{160mm}{!}{\includegraphics{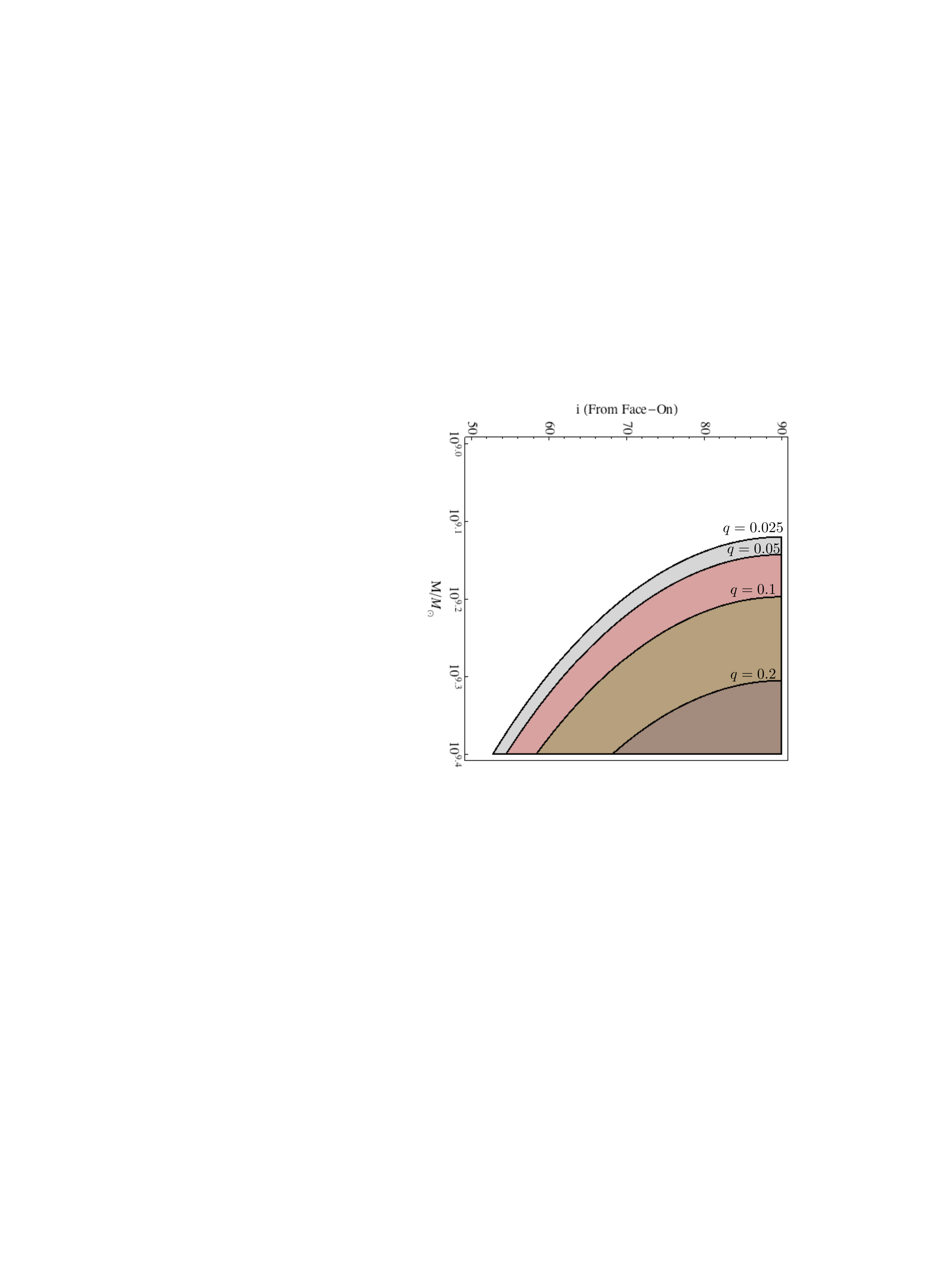}}
\end{center}
\vspace{-3in}
\noindent{\bf Extended Data Figure 2 $\mid$ Parameter combinations for
  which the combined V-band luminosity of the three--component system
  varies by the required $0.14$ mag.}  $M$ is the binary mass, $q$ is
the mass ratio, and $i$ is the orbital inclination angle. This figure
is analogous to Fig.~1, except instead of adopting a fractional
luminosity contribution $f_2$ by the secondary, the luminosities of
each of the three components are computed from a model: the primary's
luminosity is assumed to arise from an ADAF, while the secondary's
luminosity is generated by a modestly super-Eddington thin
disk. Emission from the circumbinary disk is also from a thin disk,
and is negligible except for binaries with the lowest mass ratio
$q\lsim 0.01$ (see text).

\clearpage
\newpage

\begin{center}
\resizebox{160mm}{!}{\includegraphics{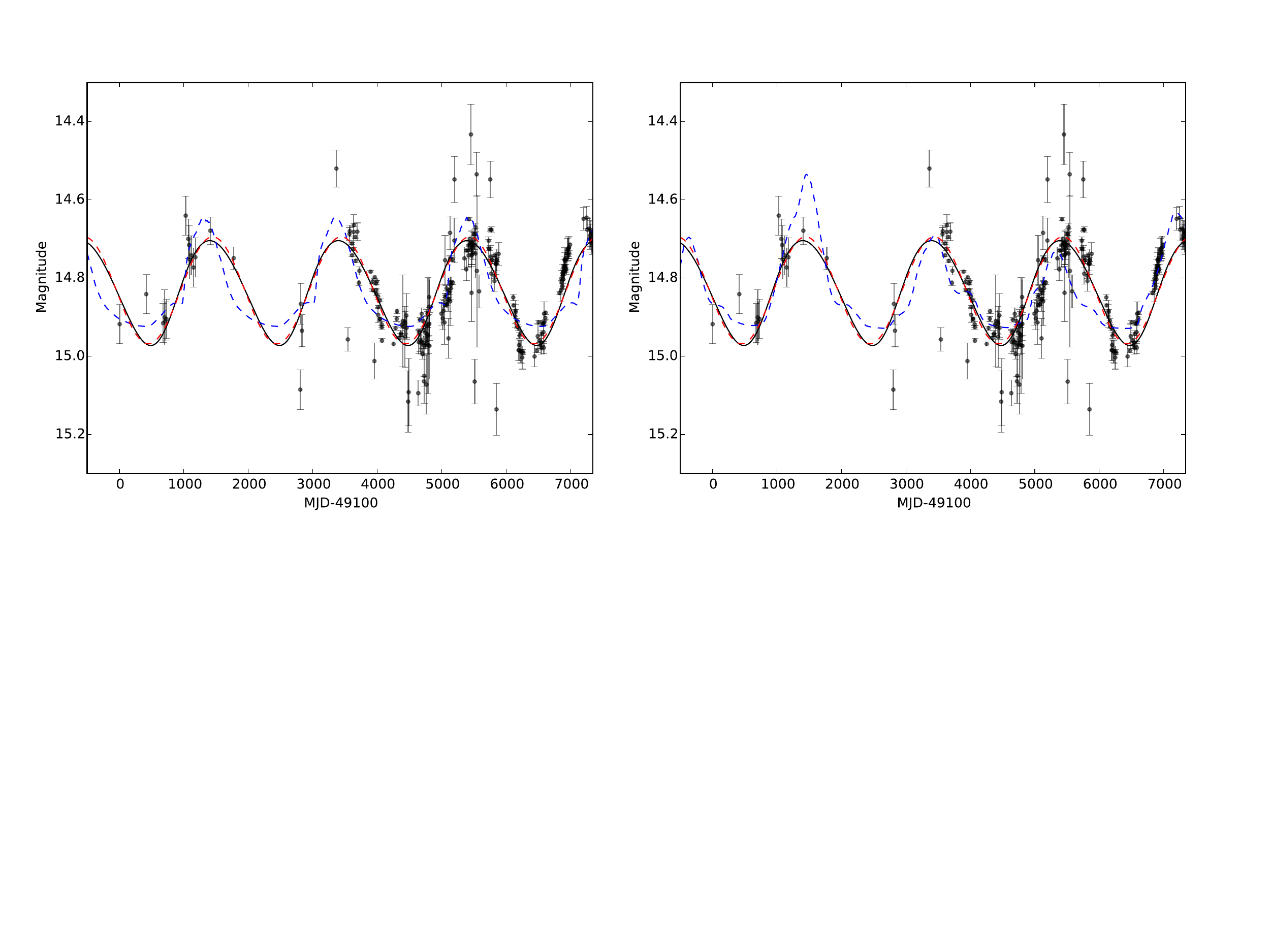}}
\end{center}
\noindent{\bf Extended Data Figure 3 $\mid$ Model fits to PG~1302-102's
  optical light curve.}  Best--fitting curves assuming relativistic
boost from a circular binary (solid black curves), a pure sinusoid
(red dotted curves) and accretion rate variability adopted from
hydrodynamical simulations\cite{Dorazio+2013} (blue dashed curves) for
a $q=0.075$ (\textbf{a}) and a $q=0.1$ (\textbf{b}) mass--ratio
binary. The grey points with $1\sigma$ errors bars show the data for
PG~1302-102\cite{Graham+2015}.

\end{document}